\documentclass[10pt, conference, letterpaper]{IEEEtran}

\usepackage{amssymb}
\usepackage{amsfonts}
\usepackage{amsmath}
\usepackage{graphicx}
\usepackage{nomencl}
\usepackage{url}

\begin{document}

\title{LoRaWAN in the Wild:\\ Measurements from The Things Network}
\author{Norbert~Blenn and Fernando~Kuipers\\
Delft University of Technology, Mekelweg 4, 2628 CD Delft, The Netherlands\\
\{N.Blenn, F.A.Kuipers\}@tudelft.nl}
\maketitle

\begin{abstract}
The Long-Range Wide-Area Network (LoRaWAN) specification was released in 2015, primarily to support the Internet-of-Things by facilitating wireless communication over long distances. Since 2015, the role-out and adoption of LoRaWAN has seen a steep growth. To the best of our knowledge, we are the first to have extensively measured, analyzed, and modeled the performance, features, and use cases of an operational LoRaWAN, namely The Things Network. Our measurement data, as presented in this paper, cover the early stages up to the production-level deployment of LoRaWAN. In particular, we analyze packet payloads, radio-signal quality, and spatio-temporal aspects, to model and estimate the performance of LoRaWAN. We also use our empirical findings in simulations to estimate the packet-loss. 
\end{abstract}

\bigskip\IEEEpeerreviewmaketitle

\section{Introduction\label{Sec_intro}}
The Long-Range Wide-Area Network (LoRaWAN\texttrademark) is a relatively new protocol in the family of Low-Power WANs (LPWANs). 
LPWANs are designed to fill the gap between (a) short-range and typically high-bandwidth networks, like Bluetooth, WiFi, and ZigBee, and (b) cellular networks, like GSM, UMTS and LTE: networks with a fairly large coverage, but also high power consumption. Since the Internet-of-Things will include many battery-operated or energy-harvesting devices, an additional requirement, which is realized by most LPWANs, is to have inexpensive low-power transceivers that are able to operate for long periods.

LoRaWAN \cite{LoRaSpec1.0}, specified by the LoRa Alliance\footnote{\url{https://www.lora-alliance.org}} in January 2015, was mainly developed to facilitate Internet-of-Things (IoT) applications \cite{Petajajarvi15,Vangelista2015,bor2016lora}. Despite its young age, the adoption of LoRaWAN has grown rapidly and its deployment by telecommunications providers suggests that it indeed is a strong contender among the set of LPWAN protocols. For example, the Dutch telecommunications operator KPN started rolling out LoRaWAN in November 2015 and within 8 months claimed to be the first operator worldwide to offer nation-wide LoRa coverage\footnote{https://corporate.kpn.com/press/press-releases/the-netherlands-has-first-nationwide-lora-network-for-internet-of-things-.htm}. The South Korean operator, SK Telecom, also reported full coverage of the country\footnote{http://www.sktelecom.com/en/press/detail.do?idx=1172} in July 2016. Furthermore, the list of members of the LoRa Alliance\footnote{https://www.lora-alliance.org/The-Alliance/Member-List} suggests that multiple telecommunications operators, electronics companies, entrepreneurs, and research institutes have adopted the technology or will do so in the near future.

In this paper, we focus on ``The Things Network'' (TTN); thethingsnetwork.org operates a real crowd-funded IoT network that can be used free of charge. LoRa gateways in the network are mainly provided by volunteers. 
The TTN site provides of map containing all gateways worldwide. Many applications make use of TTN, ranging from applications made by hobbyists to large-scale applications like UK's flood network (https://flood.network).

The outline and main contributions of this paper are as follows:  
We explain LoRa and the various means of connecting devices to a LoRaWAN network in Sec. \ref{loraexpl}, followed by a description of our large-scale measurements and data-set in Sec. \ref{Sec_crawling}. In Sec. \ref{regulations}, we explain certain regulations present in the LoRaWAN specification and their consequences in terms of the amount of traffic potentially being sent on a daily basis. In Sec. \ref{Sec:dist}, we look at estimating the distance of a device from a gateway. We develop and use an empirically-grounded simulator, in Sec. \ref{Sec_analysis}, to compute the expected packet-loss. Sec. \ref{Sec_relatedWork} describes related work. We conclude and present several guidelines in Sec. \ref{Sec_conclusion}.

\section{LoRa and LoRaWAN explained}\label{loraexpl}
\subsection{LoRa}
The LoRaWAN protocol is based on LoRa, which defines the physical (radio) layer. LoRa uses a variant of the chirp spread-spectrum (CSS PHY) modulation described in the IEEE Low-Rate Wireless Personal Area Networks (LR-WPANs) standard 802.15.4 \cite{802.15.4-2006}. 

Chirp modulation is the method of transmitting symbols by encoding them into multiple signals of increasing (up-chirp) or decreasing (down-chirp) radio frequencies. Because of the changing frequencies, chirp-modulated signals are fairly robust to multi-path interference, fading, and Doppler shifts \cite{AN1200.22}.
In chirp modulation, error-free transmission in a channel with fixed Signal-to-Noise Ratio (SNR) can be achieved by increasing the bandwidth, which is related to the number of ``chirps'' per symbol, allowing signals to be transmitted over long distances, as more information is transmitted per bit.

\subsection{The LoRaWAN protocol}
In principle, any protocol can be used on top of the LoRa protocol, but the LoRa Alliance decided to specifically develop LoRaWAN for that purpose. The reason being that other protocols, such as 6LoWPan, were expected to trigger a high amount of communication and they would ``tie'' a node to a single gateway, which would complicate the support of mobile communications. Additionally, many of the existing protocols lack(ed) security at the MAC level. LoRaWAN follows the IEEE 802.15.4 standard \cite{802.15.4-2006}, allowing seamless mobility without handovers, since any gateway will forward received LoRaWAN data frames. 

LoRaWAN \cite{LoRaSpec1.0} has three modes/classes of operation, all modes/classes referring to bi-directional communication:
\begin{itemize}
	\item[A] The basic mode, supported by \textbf{a}ll devices, is the preferred operational mode. End devices decide themselves when to send, which allows them to operate using a minimum amount of energy. After every sent data frame (up-link), a device will open two receive windows, enabling the reception of data from gateways (confirmations and down-links).
	\item[B] Additional to the functionality of class A, in class B, gateways may transmit \textbf{b}eacon frames at regular intervals, which only class B and C devices may receive.
    \item[C] In class C, devices may \textbf{c}ontinuously receive frames, except when transmitting.
\end{itemize}

\subsection{Connecting Devices to a LoRaWAN Network}\label{conndev}
\begin{figure}[ht]
	\centering
		\includegraphics[width=1.00\linewidth]{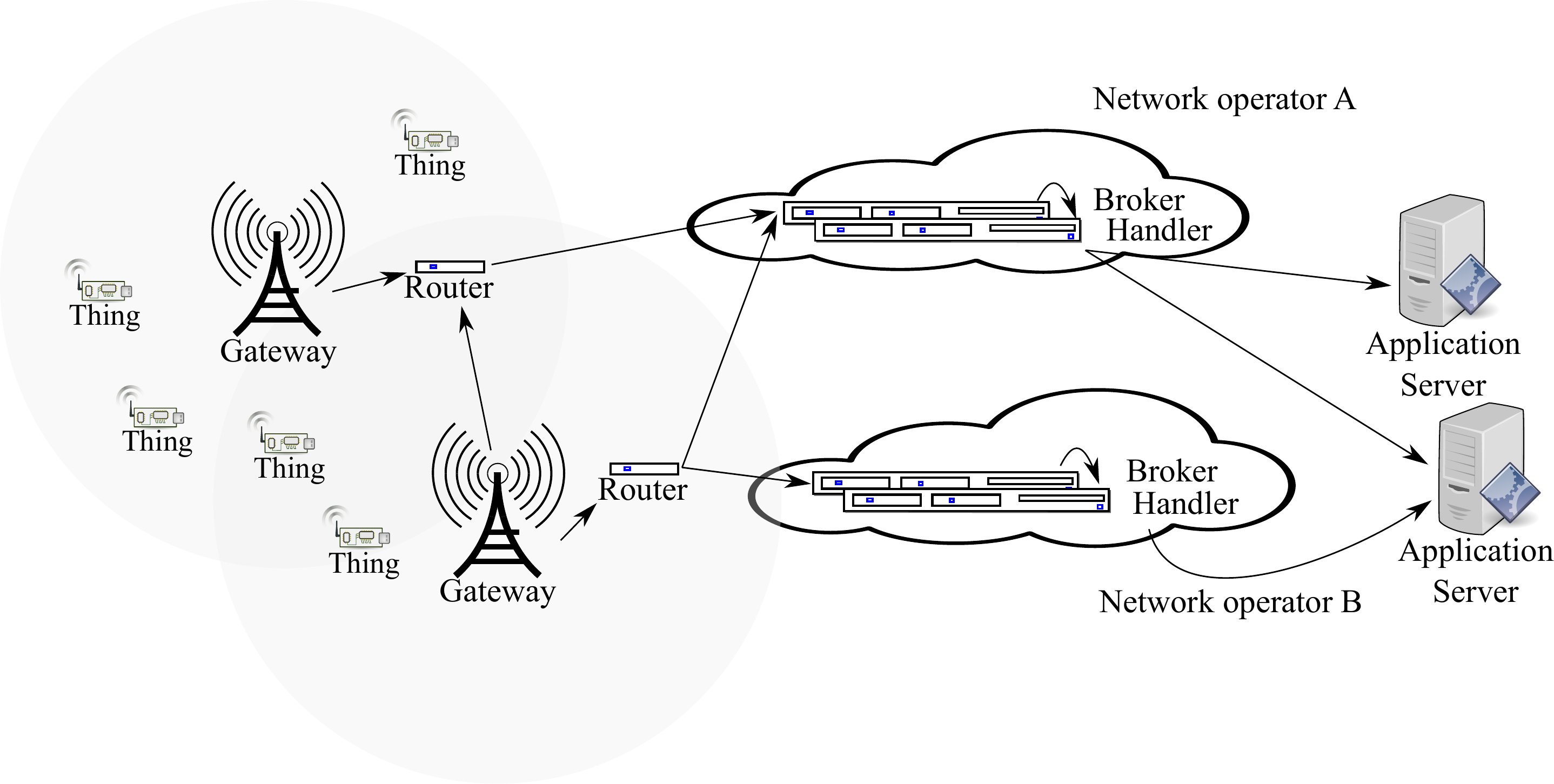}
	\caption{Overview of a typical LoRaWAN class A network. Data sent by devices are received by gateways. The gateways forward the correctly received packets to routers, which forward them to a broker based on the used network key. A broker checks the application ids of packets and forwards them to the appropriate handlers, which take care of the communication with a user's application server.}
	\label{fig:loranetwork}
\end{figure}

Fig. \ref{fig:loranetwork} illustrates a typical LoRaWAN class A network. 

A LoRaWAN network is a star-shaped access network. 
Star-shaped networks are simpler to maintain than meshed networks: Devices broadcast packets, which will be received and forwarded by gateways (typically) to a service-provider operated network. The wireless protocol (LoRa) is used for communication between gateway and device. All further communication in the backbone network is typically realized via IP networks. Applications will connect via IP to the backbone network in order to receive data from wireless nodes or to send packets to devices. 
Within the backbone network, various servers can be found, performing authentication, validation, and forwarding of packets.
Devices may connect to a network in two ways: ``Over the Air Activation'' (OTAA) and ``Activation by Personalization'' (ABP).

\subsubsection{Over the Air Activation (OTAA)}

A device needs to be equipped with a DevEUI, which is a 64-bit globally-unique identifier of the device, an owner-unique AppEUI of 64 bits, which identifies the application the device wants to connect to, and a 128-bit AppKey. The AppKey, obtained from the network operator after successful registration of a device, is used to sign an initial join-request, including the DevEUI, AppEUI, and a randomly generated two-byte DevNonce, which is signed by a Message Integrity Code (MIC).
A server that validates the MIC may respond, within the time the receive windows of the device are open, with a new nonce (AppNonce), a 128-bit AppSKey (application session key), a 128-bit NwkSKey (network session key), a device address (DevAddr), RF delays (RxDelay), as well as channels to use (CFList), in a message signed by a MIC.

\subsubsection{Activation by Personalization (ABP)}

One may also skip OTAA and directly supply devices with a DevAddr, NwkSKey, and AppSKey to send packets. In that case, a device is typically manually registered at the service operator to obtain the keys directly. 

\subsubsection{Default Activation for Generic Devices}

A device that uses default keys is called a generic device. A generic device does not need to register its device id nor does it need an associated application, since all network operators should support the default keys. Packets sent by generic devices often are not encrypted. Generic devices use ABP with globally-known NwkSKey and AppSKey.
Security is only partially available, namely when individual AppSKeys are used to encrypt data. However, meta-data of transmissions, like the time at which a packet was sent, the length of that packet, the DevAddr, signal strength, SNR, as well as gateway information, are visible to the public. By convention, the AppSKey equals the NwkSKey, in which case everyone is able to decode the data.

Many network operators, like ``The Things Network'', have supported generic keys\footnote{Default Semtech keys: AppSKey = NwkSKey: 2B7E151628AED2A6ABF7158809CF4F3C.}, but this might change due to server-sided routing issues. For example, when a user registers a device on a network using the mentioned generic keys, it is unclear for the operator whether the packets sent by the registered device should be routed to a public interface, or to the application of the user.

\subsection{Data Encryption}
After a device connects to a network, packets are encrypted using a user-supplied key. LoRaWAN uses AES128 for encryption and adds a frame counter to the packets, whereas the application payload is encrypted by the AppSKey and the whole packet, including the frame counter and the DevAddr, is signed by the NwkSKey. As the NwkSkey is only known to the node and the network server, the integrity of a packet can only be verified within the network where the device is registered. In this case, the server checks the MIC of the received frame against the corresponding key in its key database. 

Once the message integrity has been verified, the packet will be forwarded to the user's application server or to an endpoint delivering packets to the application. Only the owner of the AppSKey can then decrypt the packet's payload.

Keys need to be stored in at least two locations: the network server and the node's memory. A user therefore needs to take the necessary steps to secure access to the nodes. If an attacker would obtain keys from a node, it would be possible to intercept or inject falsified/malicious traffic. Each device should therefore use different keys to avoid that the theft of data from one node compromises all other nodes.

\section{Data Collection and Analysis\label{Sec_crawling}}
In this section, we present and analyze our measurement data.

\subsection{Data Collection}
All large-scale LoRaWANs currently deployed operate in class A. In this paper, we therefore focus on class A networks.

We have captured all data received by the gateways from ``The Things Network'' between December 2015 until July 2016,  which were sent by nodes using ABP (activation by personalization) with the generic NwkSkey: 2B7E151628AED2A6ABF7158809CF4F3C. Additionally, between May and July 2016, we have obtained 23.5 million gateway status updates from gateways in the network.

Table \ref{Tab:dataset} describes our data-set, which was obtained through the API of ``The Things Network.'' 
The data comprise two perspectives of all frames\footnote{We use the terms ``frames'' and ``packets'' interchangeably.} sent by devices using the previously mentioned generic key. 
On the one hand, raw data of all frames received by gateways and, on the other hand, information sent every 30 s by gateways containing aggregated numbers of received and sent frames. 

\begin{table}[ht]
\centering
\begin{tabular}{l c}
frames received by gateways & 17,467,312 \\
unique frames received by gateways & 16,228,814 \\
unique device ids & 1,618\\
gateways & 691\\
size of the data-set & 9.4 GB\\
\end{tabular}
\caption{Statistics of the collected data-set.}
\label{Tab:dataset}
\end{table}

Frames sent by LoRa devices are potentially received by multiple gateways. The number of received frames in the data-set is therefore larger than the number of unique frames received. 94.8\% of the unique frames were received by one gateway, 3.7\% by two gateways, and 1.1\% by three gateways. The highest number of gateways that received one frame was 31. That particular frame originated from a LoRaMote\footnote{A LoRaMote\texttrademark is a device developed to test LoRaWAN connectivity, equipped with a sensor for atmospheric pressure, a temperature sensor, an accelerometer, a GPS receiver, and a LoRa transceiver.} that transmitted a packet from an altitude of 1.4 km.

\begin{figure}[ht]
	\centering
		\includegraphics[width=1.00\linewidth]{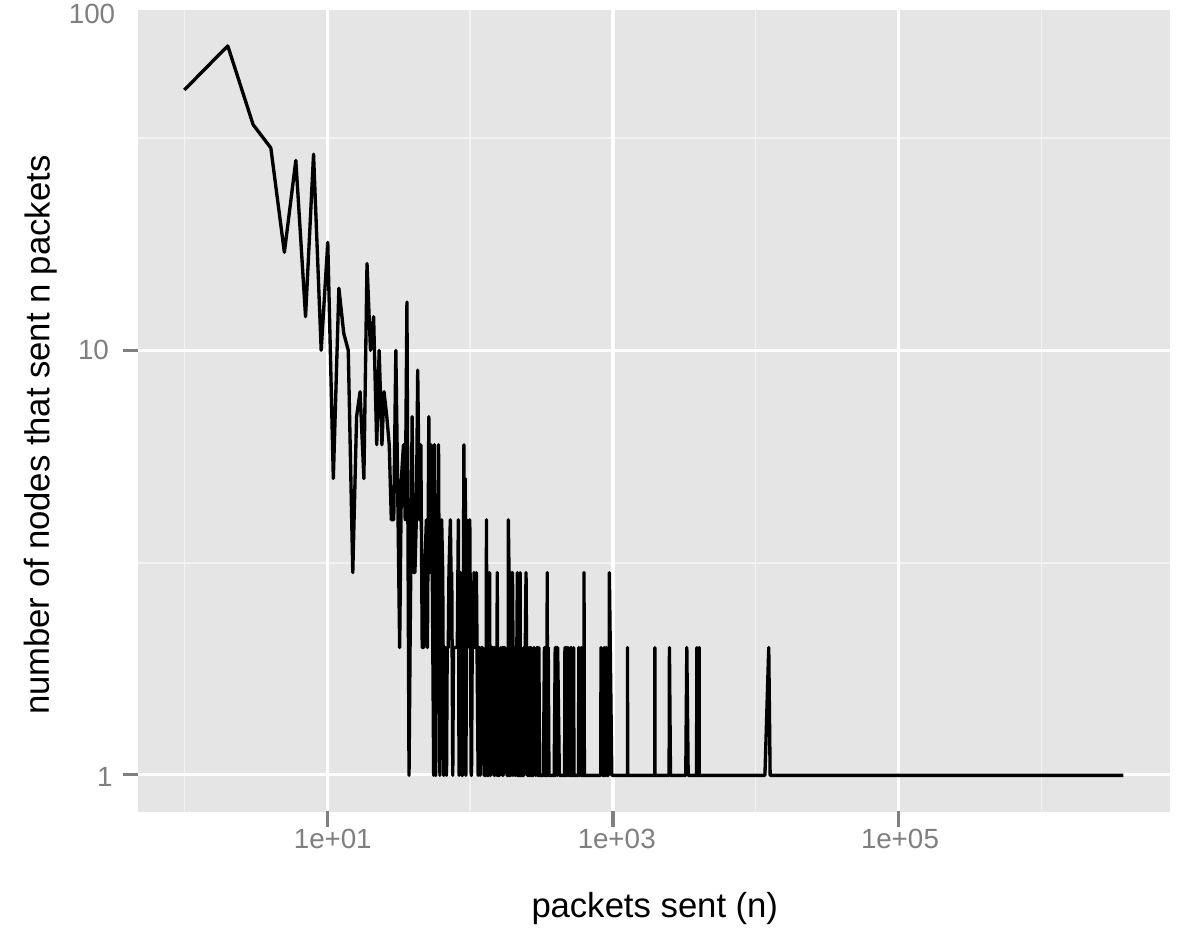}
	\caption{Histogram of the number of devices that sent $n$ packets.}
	\label{fig:histnodes2freq}
\end{figure}

Fig. \ref{fig:histnodes2freq} shows that 
the number of unique frames, sent by individual devices, approaches a power law, which indicates a very skewed use pattern of devices. Very few devices sent more than 1,000 packets, whereas a high number of devices sent less than 50 packets. 

\subsection{Signal quality}
\begin{figure}[ht]
	\centering
		\includegraphics[width=1.00\linewidth]{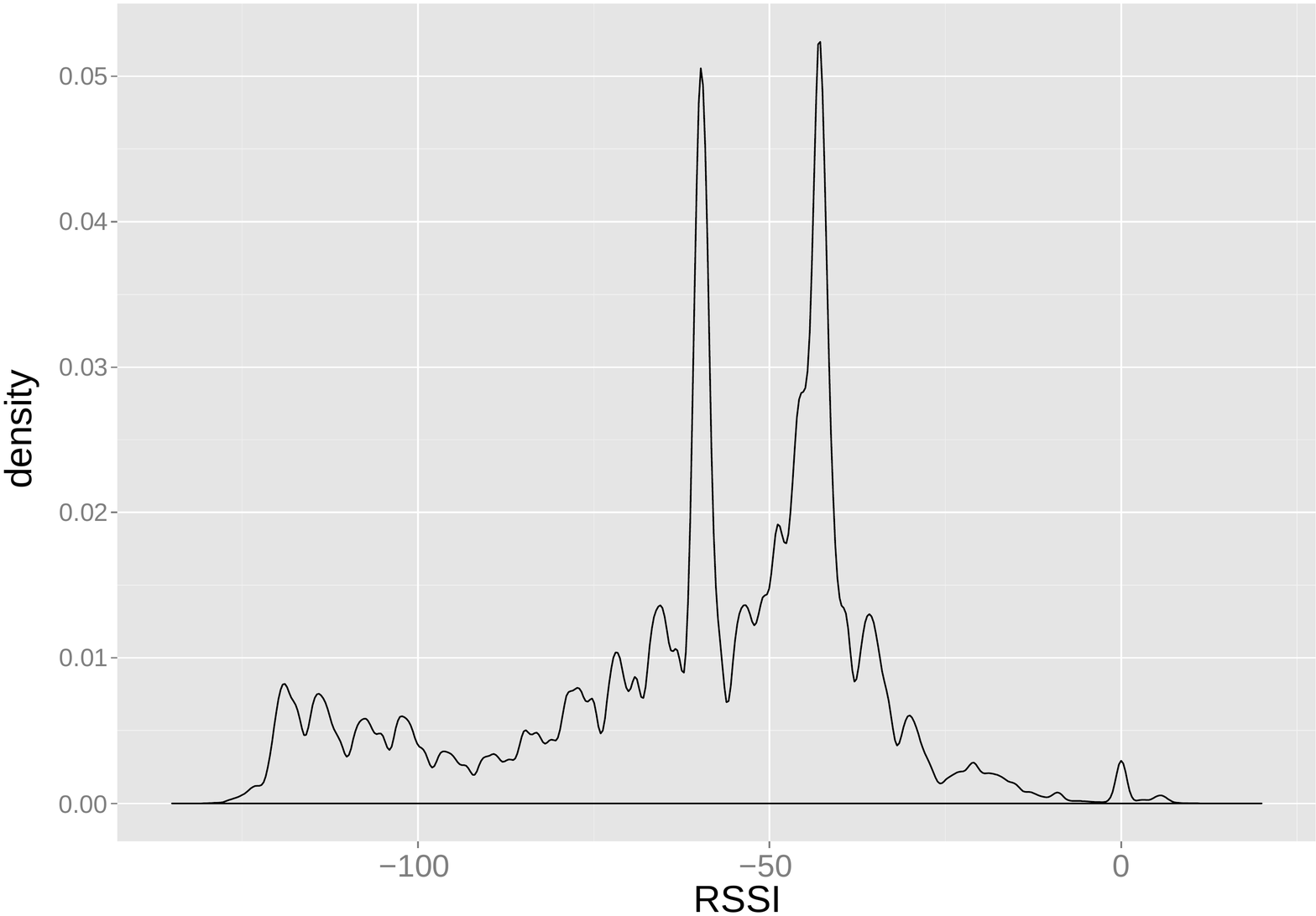}
	\caption{Probability-density function of RSSI values of all received frames, as reported by gateways.}
	\label{fig:histrssi}
\end{figure}

\begin{figure}[ht]
	\centering
		\includegraphics[width=1.00\linewidth]{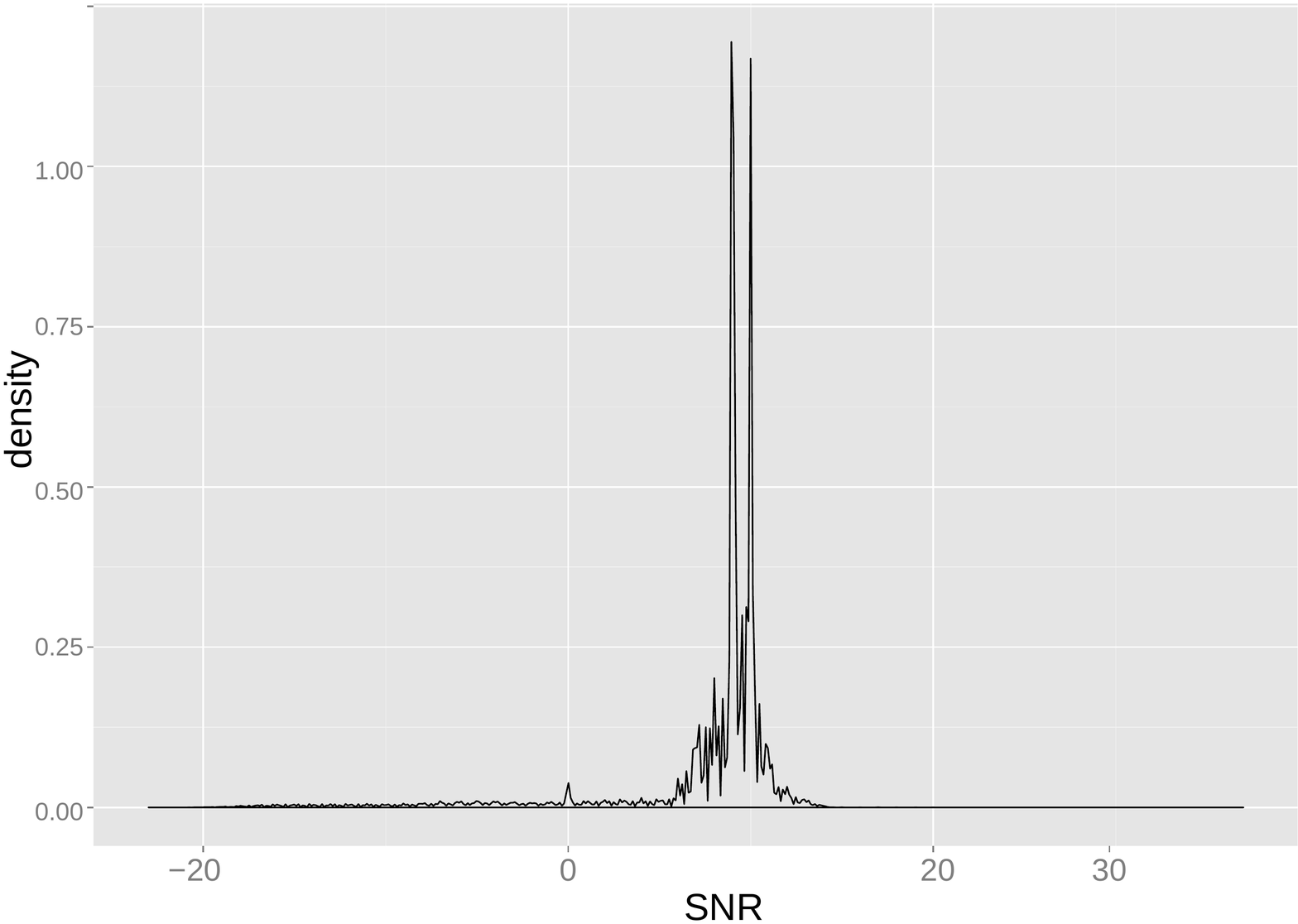}
	\caption{Probability-density function of SNR values in dB of all received frames, as reported by gateways.}
	\label{fig:histsnr}
\end{figure}

Figures \ref{fig:histrssi} and 
\ref{fig:histsnr} depict the probability-density function for the RSSI and 
SNR values, as captured for all frames in our data-set.

The figures illustrate relatively small RSSI values and 
positive SNR values, which suggests that the majority of received data was sent by devices close to gateways.  

\subsection{Payload analysis}

\begin{figure}[ht]
	\centering
		\includegraphics[width=1.00\linewidth]{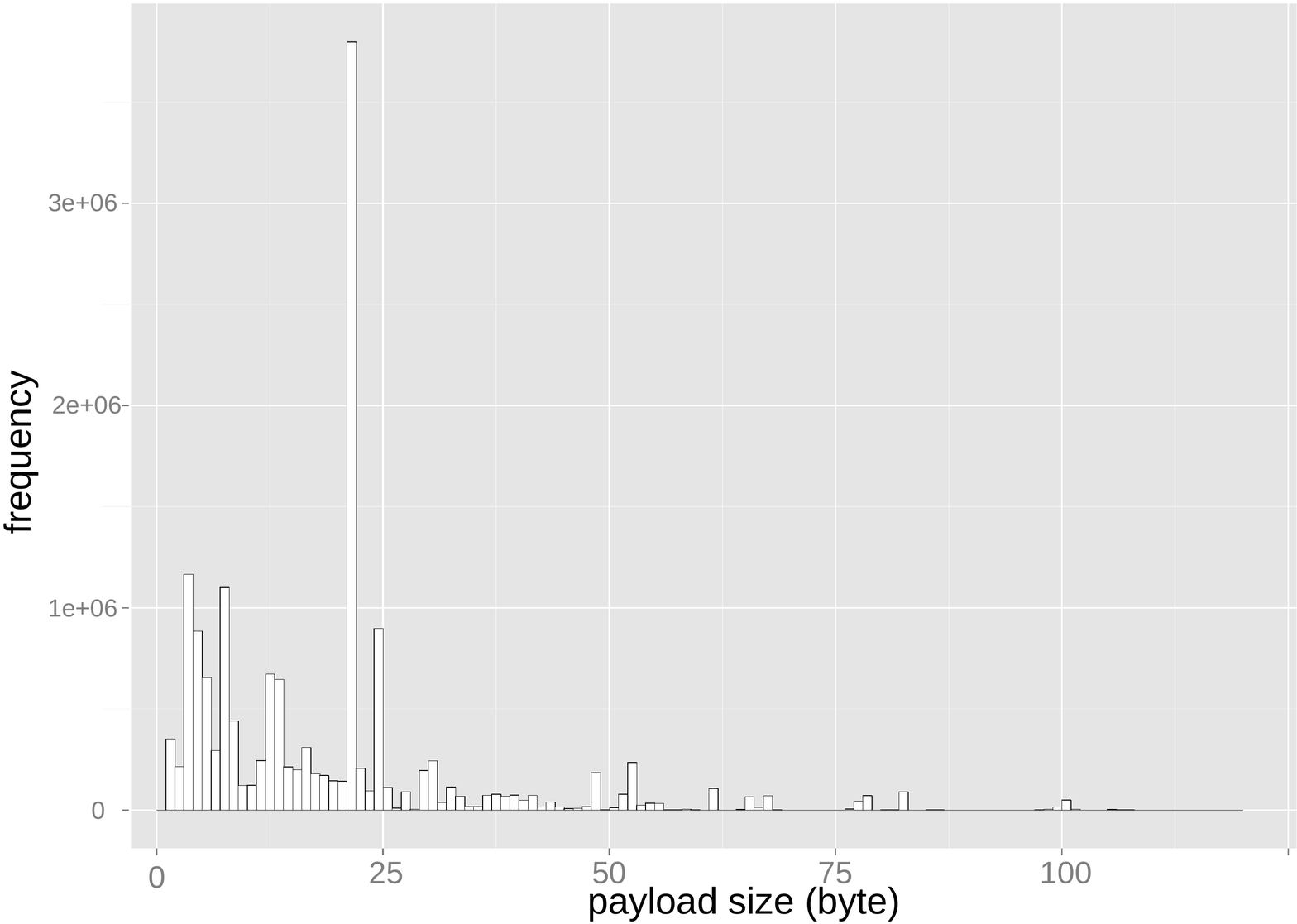}
	\caption{Histogram of payload size in bytes.}
	\label{fig:histsizepayload}
\end{figure}

We have analyzed the payloads of all unique frames in our data-set. 
As depicted in Fig. \ref{fig:histsizepayload}, 93.7\% of the captured payloads are smaller than 50 bytes and 50\% of the payloads are even smaller than 19 bytes, whereas the average payload size is 18 bytes. 

Out of all unique frames, 97,888 were sent by LoRaMotes. 
The payload of the remaining 16.1 million frames contained 10,104,330 human-readable strings and roughly 6 million payloads that we could not decipher/decode. Table \ref{readable:class} gives a classification of the human-readable strings.

\begin{table}[ht]
\centering
\begin{tabular}{l l}
comma separated decimals & ca. 5.54 million \\
temperature readings & ca. 1.3 million\\
various other strings & ca. 1 million \\
string: foo... & 974,634\\
string: hello & 733,724\\
humidity measurements: & 666,609\\
GPS locations & 320,391\\
battery level & 140,450\\
light sensor (brightness) & ca. 45,000\\
string: test & 42,336\\
distance measures: & ca. 2,500 \\
string: coffee & 172
\end{tabular}
\caption{Overview of human-readable payloads.}
\label{readable:class}
\end{table}

As multiple sensor measurements, like battery and temperature values, are often combined in one frame, the counts in Table \ref{readable:class} sum up to a larger number than 10.1 million. Nonetheless, Table \ref{readable:class} provides a good indication of the use of LoRaWAN devices on The Things Network (for our measurement time frame).

\subsection{Spreading Factor}

\begin{figure}[ht]
	\centering
		\includegraphics[width=1.00\linewidth]{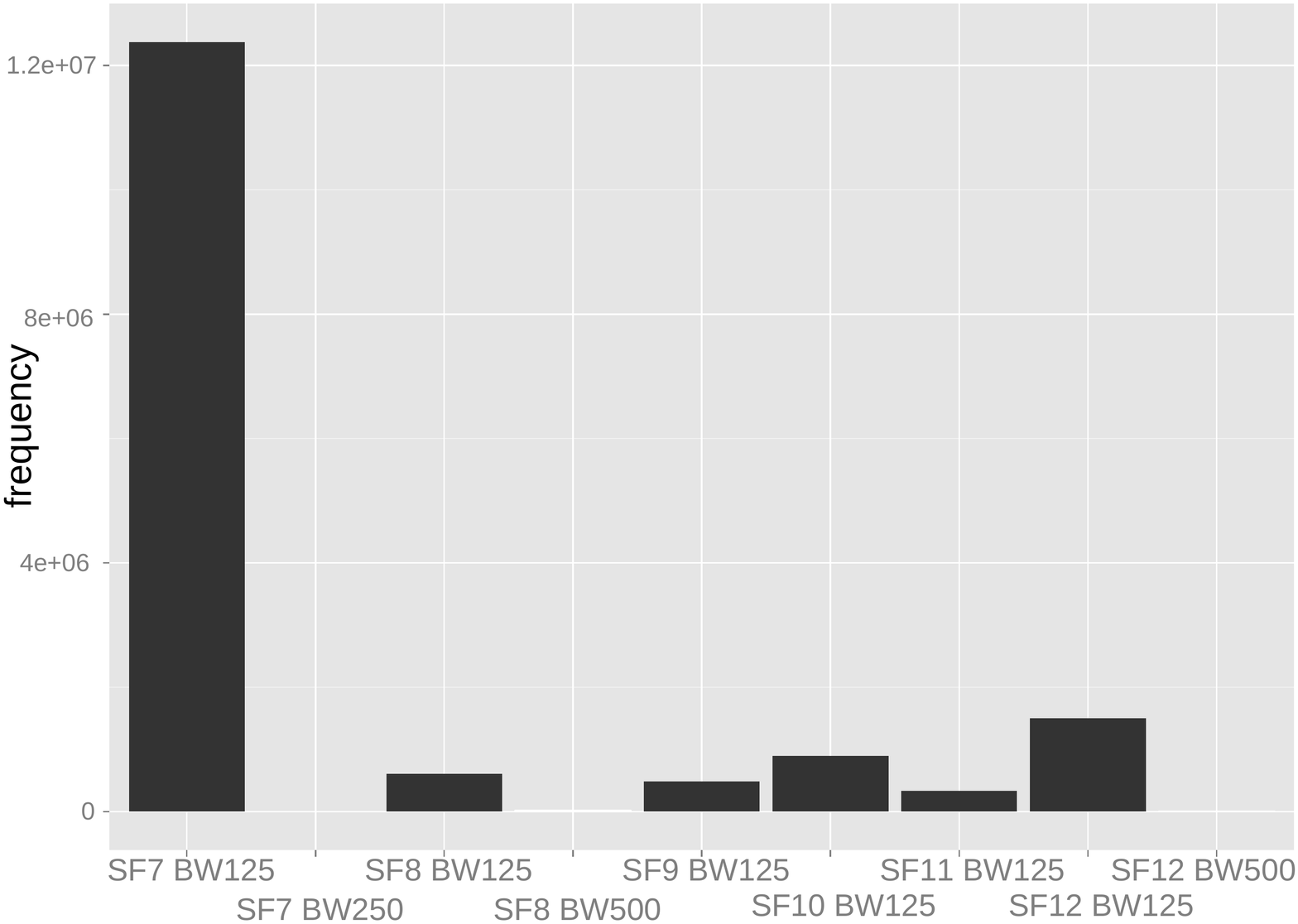}
	\caption{Histogram of the used spreading factors (SF).}
	\label{fig:histSF}
\end{figure}

Spreading factors range from SF7 to SF12 and denote the number of chirps used to encode a bit. For example, SF7 encodes each bit into 128 ($2^7$) chirps, whereas in SF12 each bit is encoded into 4096 ($2^{12}$) chirps. A higher chirp rate enables a better reconstruction of the received signal, but also stretches the duration needed to send a bit.\\
Fig. \ref{fig:histSF} displays the spreading factors (SF) used to send unique frames in our data-set. Most devices used SF7 125kHz, which is the default setting for devices using the open-source LoRa implementation provided by IBM\footnote{https://www.research.ibm.com/labs/zurich/ics/lrsc/lmic.html} and devices manufactured by Microchip\texttrademark.

\begin{figure}[ht]
	\centering
		\includegraphics[width=1.00\linewidth]{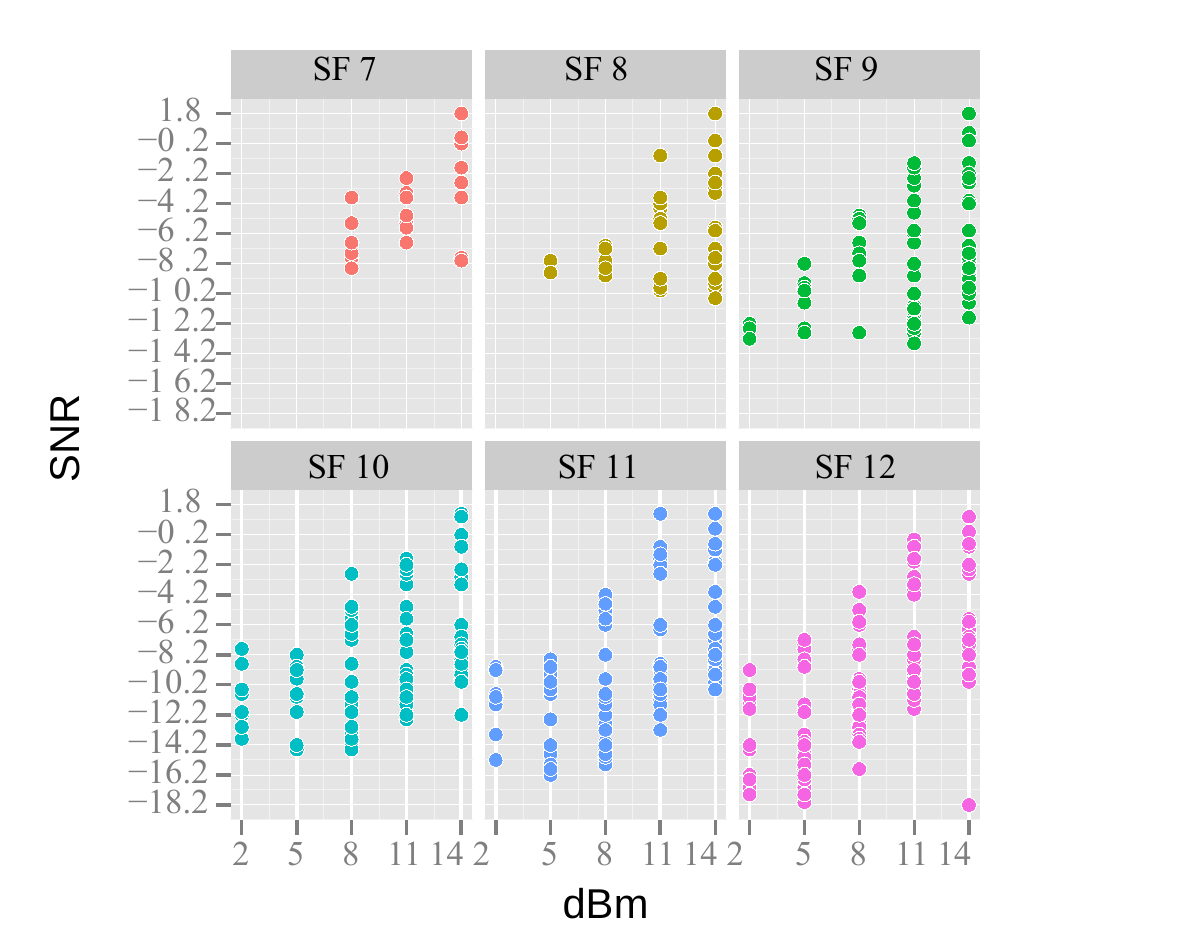}
	\caption{Measured SNR under varying spreading factors and transmit power.}
	\label{fig:dBm2SF2snr}
\end{figure}

To analyze the influence of different spreading factors and power utilized in transmitting frames, we employed our own LoRa node. Our tests were conducted from one location, by sending packets having the same (one character) payload, at varying spreading factors and power settings. In total, we used all 6 spreading factors (0-5 in Table \ref{tab:datarates}) and 5 power settings (2, 5, 8, 11, 14 dBm) and transmitted 20 packets per configuration. 
The results of these tests are shown in Fig. \ref{fig:dBm2SF2snr}.

The different sub-figures in Fig. \ref{fig:dBm2SF2snr} depict which spreading factors were used to transmit a frame.
One may observe that using higher spreading factors and increasing the transmit power, reduce the observed packet-loss. For example, compare SF7, where no frames are received for output powers below 8 dBm, to SF12, for which these frames are received, even at 2 dBm transmit power. In contrast, sending packets at high power (14 dBm) results, in this example, in frames being received with SF7.

\section{Frequency Usage and Regulation}
\label{regulations}
According to the frequencies used by nodes in our data-set, we observe that 89.4\% of the packets were sent within the 868 MHz band, 10.5\% in the 902-928 MHz band, and 0.1\% within the 433 MHz band. According to the LoRa specification, Europe uses the 863-870 Mhz and the 433 MHz ISM bands\footnote{Industrial, Scientific, Medical (ISM) bands are regulated, but free to use by certain domains, see ETSI \cite{EN300.200}.}, the USA uses the 902-928 MHz band, and China uses the 779-787 MHz band. Our results therefore suggest that most devices in our data-set were located within Europe and the USA.

For Europe, where The Things Network is indeed most prominent, LoRaWAN specifies 3 channels (868.10, 868.30, 868.50 MHz of 125 kHz bandwidth), a data-rate between DR0 and DR5 (see Table \ref{tab:datarates}), a duty-cycle of $<1\%$, and a default radiated transmit power of 14 dBm \cite{LoRaSpec1.0}.

\begin{table}
\centering
\begin{tabular}{c c c}
Data-rate & Configuration & Indicative physical bit rate [bit/s]\\
\hline\\
0	& LoRa: SF12 / 125 kHz & 250 \\
1	& LoRa: SF11 / 125 kHz & 440 \\
2	& LoRa: SF10 / 125 kHz & 980 \\
3	& LoRa: SF9 / 125 kHz & 1760 \\
4	& LoRa: SF8 / 125 kHz & 3125 \\
5 	& LoRa: SF7 / 125 kHz & 5470 \\
6 	& LoRa: SF7 / 250 kHz & 11000\\
7 	& FSK: 50 kbps 440 & 50000\\
8..15 & RFU & \\
\end{tabular}
\caption{EU 863-870 MHz data-rates \cite{LoRaSpec1.0}.}
\label{tab:datarates}
\end{table}

\begin{figure}[ht]
	\centering
		\includegraphics[width=1.00\linewidth]{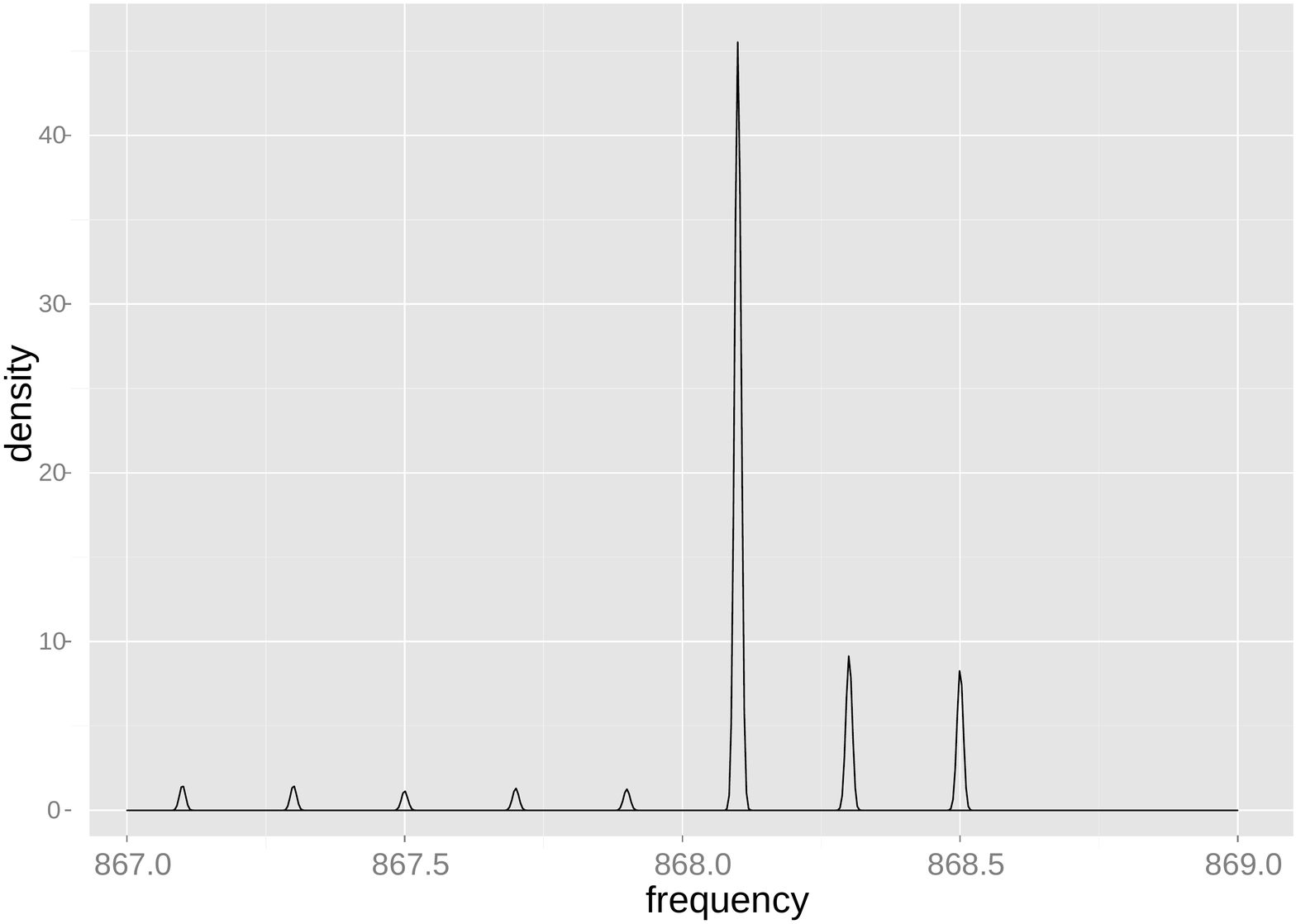}
	\caption{Histogram of the usage of the 863-870 MHz band.}
	\label{fig:histfreq}
\end{figure}

Fig. \ref{fig:histfreq} depicts the frequencies in the 863-870 MHz band that were encountered in our data-set. The 3 compulsory frequencies of 868.1, 868.3, and 868.5 MHz were indeed used most. The other frequencies (867.1, 867.3, 867.5, 867.7, and 867.9 MHz) are supported by individual network operators.

The time it takes to transmit a frame from node to gateway is calculated via the following set of equations \cite{AN1200.13}:
\begin{equation}
T_{sym} = \frac{2^{SF}}{BW} \label{tsym}
\end{equation}
\begin{equation}
T_{preamble} = \left(n_{preamble}+4.25\right) T_{sym} \label{tpre}
\end{equation}
\begin{equation}
\beta = \left\lceil\frac{8PL-4SF+28+16-20H}{4(SF-2DE)}\right\rceil \label{beta}
\end{equation}
\begin{equation}
PL_{sym} = \max\left(\beta\left(CR+4\right),0\right)+8 \label{nsym}
\end{equation}
\begin{equation}
T_{frame} = T_{preamble} + PL_{sym} \times T_{sym} \label{tpac}
\end{equation}
where $PL$ denotes the size of the payload in bytes, $SF$ the spreading factor, $n_{preamble}$ is the number of preamble symbols, $H = 0$ if the header is enabled and 1 if it is not (for LoRaWAN the header is always present, whereas for pure LoRa one may create frames without header information), $DE = 1$ if low-data rate optimization is enabled and 0 otherwise (this optimization is active for SF11 and SF12 to account for drift of the crystal reference oscillator during long transmissions).
The coding rate ($CR$) relates to Forward Error Correction. Higher values imply better reconstruction of noisy signals.

Using the time to transmit one symbol, Eq. (\ref{tsym}), the number of symbols in the payload, Eq. (\ref{nsym}), and the size of the preamble (8 symbols in LoRaWAN), one may calculate the time needed to transmit a whole frame by Eq. (\ref{tpac}).

\subsection{Limits}

\begin{figure}[ht]
	\centering
		\includegraphics[width=1.00\linewidth]{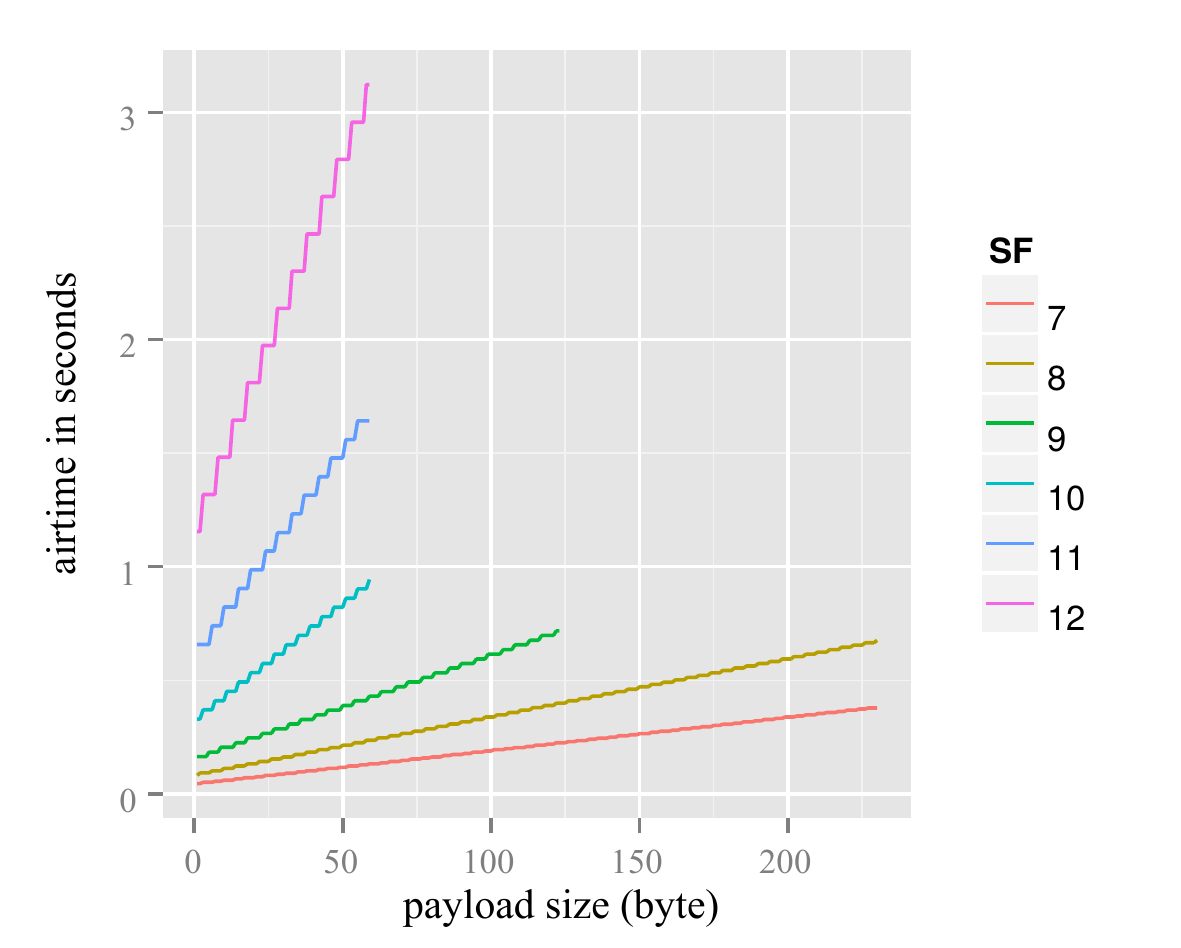}
	\caption{Airtime for different SF and payloads with BW = 125kHz and coding rate $\frac{4}{5}$ according to Eq. (\ref{tpac}).}
	\label{fig:airtime}
\end{figure}

We have calculated several theoretical limits, assuming a node follows the specification. As depicted in Fig. \ref{fig:airtime}, the airtime (time to transmit one frame) linearly increases with the size of the payload. According to the specification, the maximum payload for frames using SF7 and SF8 is 230 bytes, 123 for SF9, and 59 for SF10-12.

\begin{figure}[ht]
	\centering
		\includegraphics[width=1.00\linewidth]{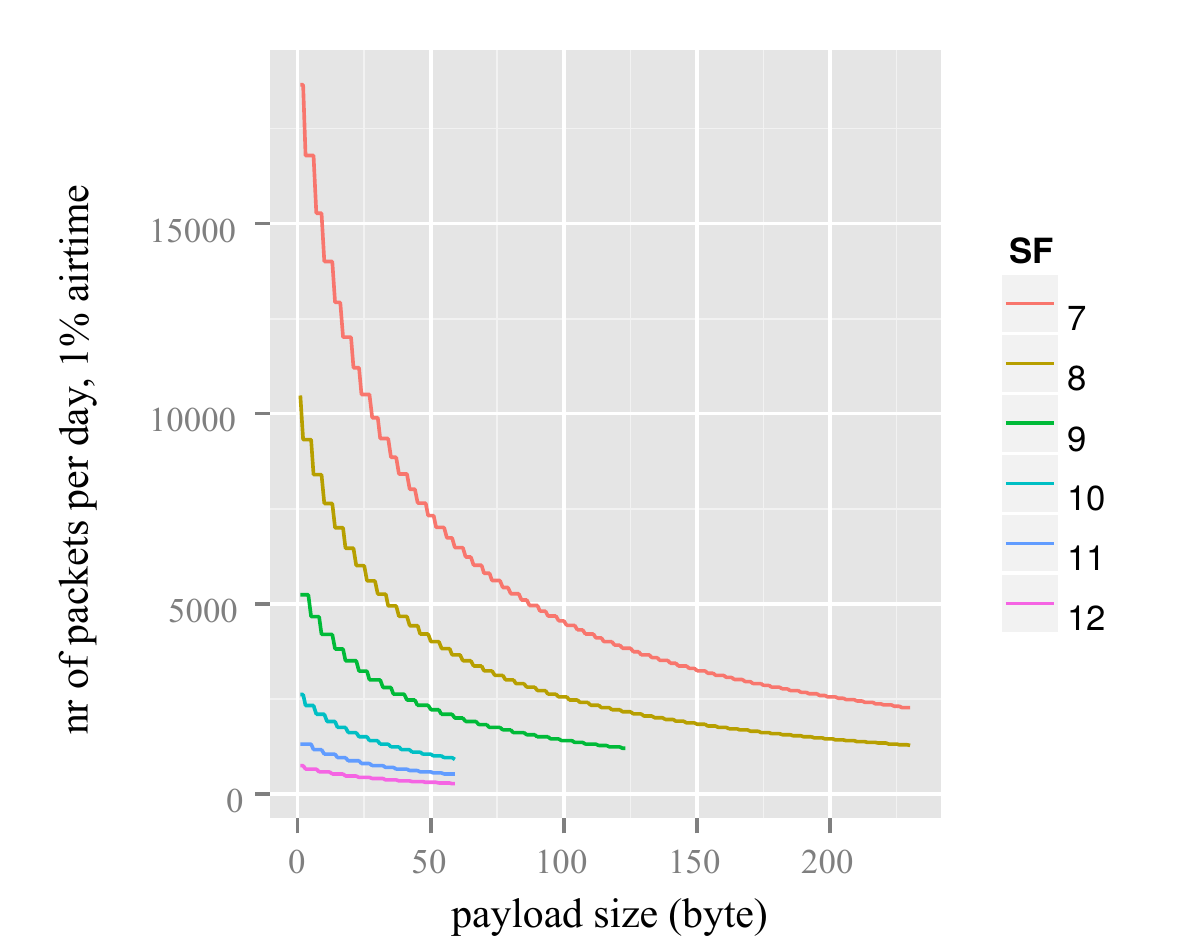}
	\caption{Number of frames that can be sent per day for different SF and payloads with BW = 125kHz and coding rate $\frac{4}{5}$.}
	\label{fig:ppd}
\end{figure}

\begin{figure}[ht]
	\centering
		\includegraphics[width=1.00\linewidth]{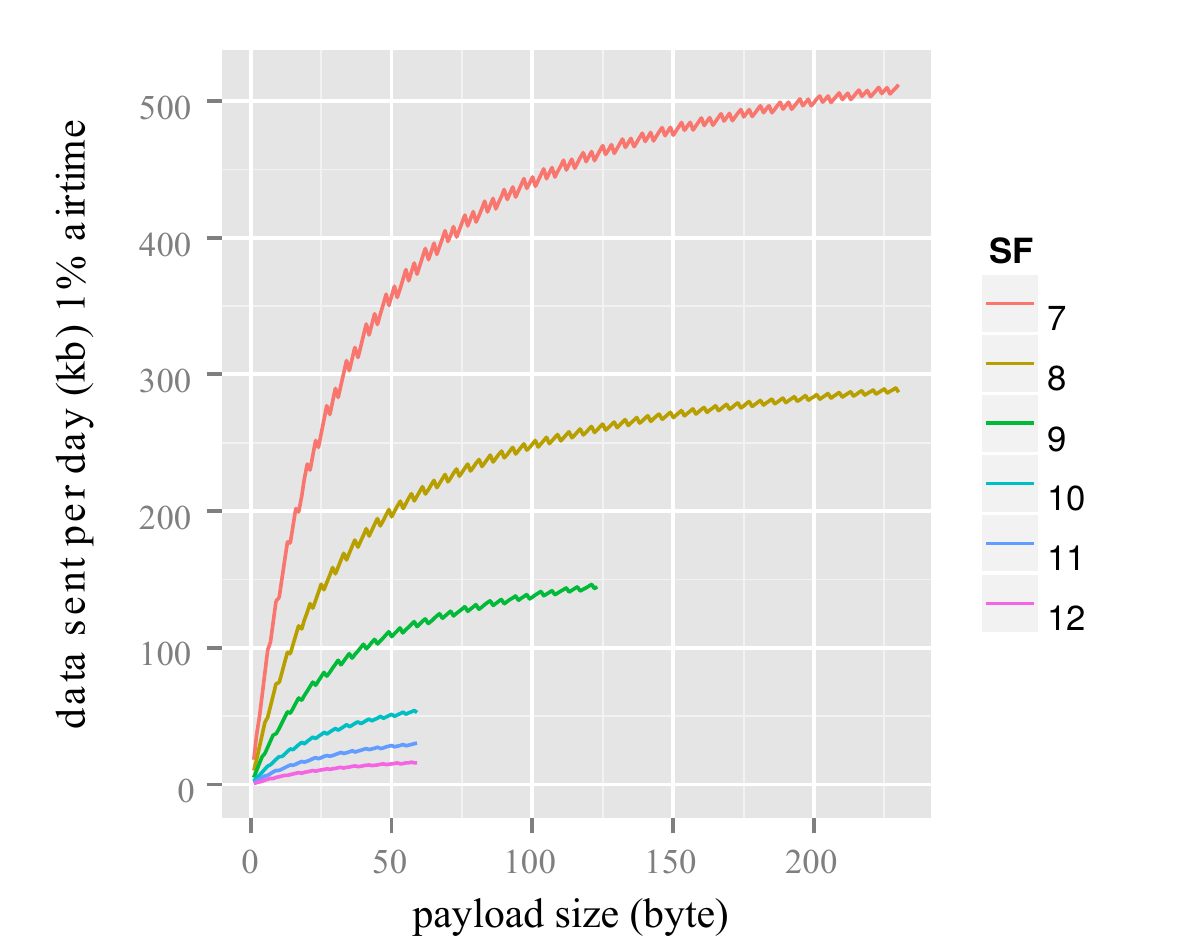}
	\caption{Amount of data that can be sent per day for different SF and payloads with BW = 125kHz and coding rate $\frac{4}{5}$.}
	\label{fig:dpd}
\end{figure}

Figures \ref{fig:ppd} and \ref{fig:dpd} depict the effect of different airtime values on the maximum number of frames a node may transmit per day in one channel.

These estimates are based on the assumption that a device is sending unconfirmed frames and does not activate via OTAA. It is possible to send confirmed frames; frames which are acknowledged by the gateway(s). If the response, however, is not received, a node would re-transmit that frame for a pre-configured number of times, which decreases the number of frames and data that can be transmitted per day.

\section{Distance estimation}\label{Sec:dist}

From the used data-set, we were able to obtain some GPS locations, sent either by LoRaMotes or by individuals sending text containing a latitude, longitude tuple. The total number of captured frames containing GPS coordinates is 320,391.

Comparing the distance from these known locations to the locations of gateways receiving frames, allows us to compare distance estimations to actual measurements.

To estimate the distance of nodes to gateways, and in absence of a specific LoRa path-loss equation, we employ the often-used approximation based on the free-space path-loss equation \cite{erceg1999empirically}, as shown in Eq. (\ref{distest}).

\begin{equation}
\label{distest}
	d = 10^{27.55 - (20 \log_{10}(f)) + |s|) / 20.0}
\end{equation}

\begin{figure}[ht]
	\centering
		\includegraphics[width=1.00\linewidth]{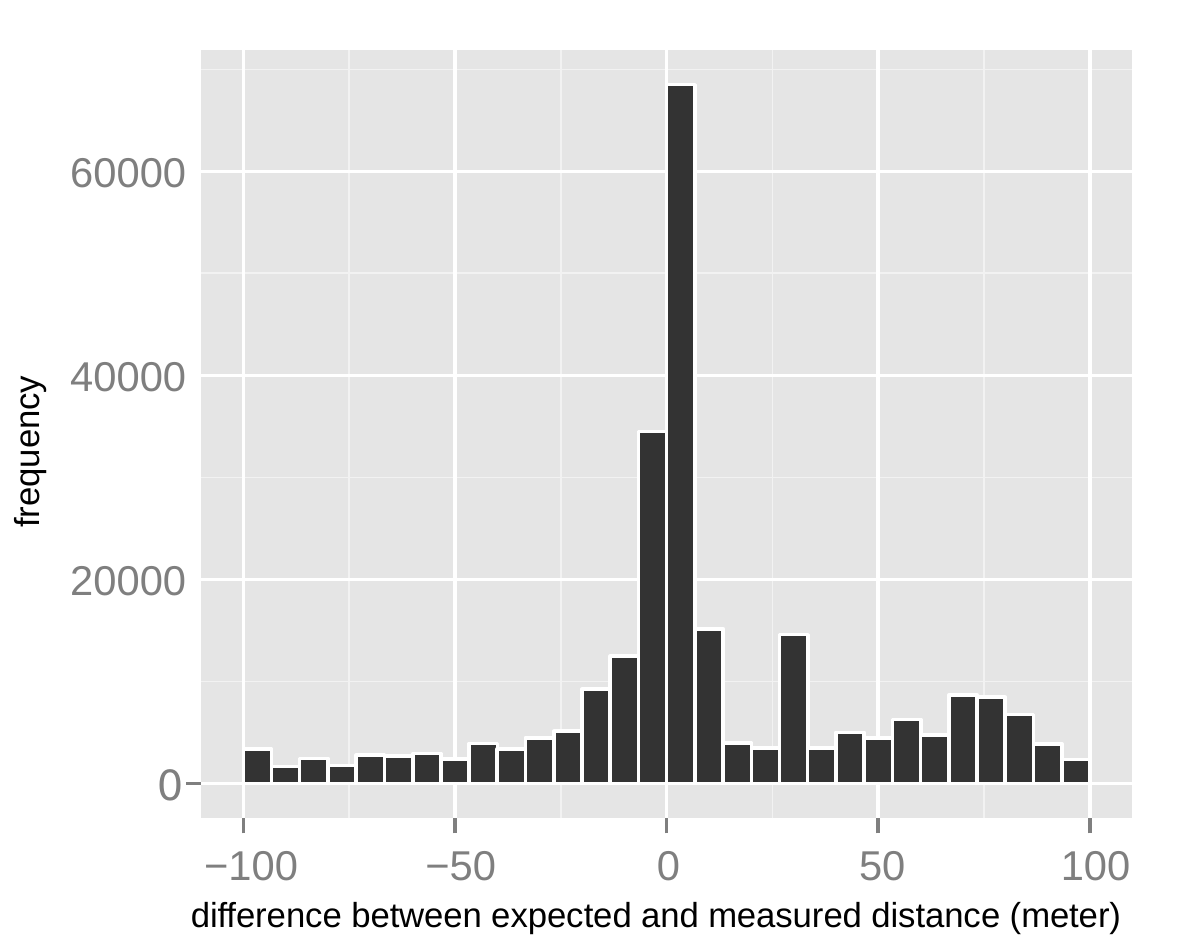}
	\caption{Histogram depicting the differences between measured and estimated distances, between LoRa nodes and gateways, for a range of -100 to 100 meters.}
	\label{fig:dist_m}
\end{figure}

Here $f$ denotes the frequency in MHz, $s$ the received signal level in dB and $d$ the expected distance in meters.
Fig. \ref{fig:dist_m} depicts a histogram of the differences between measured and estimated distances.

Our results look quite promising, as the majority of estimations match the measurements and hardly any difference exceeded 100 m. The observable high number of positive differences denote the difficulties of using Eq. (\ref{distest}) in urban environments. In such scenarios, the measured distance is less than the estimated one, as the signal is dampened by buildings or other urban structures.

\section{Simulations \label{Sec_analysis}}
To further explore the practical limitations of LoRaWAN, we have developed a simulator that takes some observations from our empirical data as input. 

\subsection{Packet loss}

In order to understand under which circumstances packets might get lost, apart from being too far from any receiving gateway, we conducted a physical experiment using two nodes at the same location. The results are shown in Table \ref{tab:exploss}.

\begin{table}
\centering
\begin{tabular}{c c c | c c c | c}
 	& node 1& 		 &	   & node 2& 		& \\
SF  & dBm  & channel & SF  & dBm  & channel & packet forwarded\\
\hline\\
11  & 14  & 6 & 11  & 14  & 6 & none\\
11  & 11  & 6 & 11  & 14  & 6 & none\\
11  & 8  & 6 & 11  & 14  & 6 & none\\
10  & 14  & 6 & 11  & 14  & 6 & both\\
11  & 14  & 1 & 11  & 14  & 6 & both\\
    &     &   &     &     &   & \\
\end{tabular}
\caption{Packet-loss under different settings.}
\label{tab:exploss}
\end{table}

For each configuration listed in Table \ref{tab:exploss}, both devices sent 120 packets with the same payload and at the same time. The table illustrates that the arrival of multiple frames using the same spreading factor and channel at the same time, will lead to collisions and hence packet loss. A similar experiment was conducted by Bor \textit{et al.} \cite{bor2016lora}, in which the authors found that, in plain LoRa, the stronger signal is recoverable. However, in our tests, using two Microchip RN2483 transceivers sending LoRaWAN packets to a Lorank 8 gateway\footnote{http://www.ideetron.nl/lora}, we obtained different results. As the gateway checks the integrity of the received frames, corrupted data is not forwarded, and therefore attributes to packet loss.

Our observations allow us to employ a simulation to estimate the number of collisions, at a single gateway, under arbitrary assumptions. 
We consider $n$ packets, each having a payload of 1 byte,  where every packet:
\begin{itemize}
	\item was sent with a SF selected randomly from a uniform distribution of all spreading factors (7-12), and
	\item was sent on a channel selected at random from a uniform distribution out of the 3 EU channels defined in the LoRaWAN specification.
\end{itemize}

\begin{figure}[ht]
	\centering
		\includegraphics[width=1.00\linewidth]{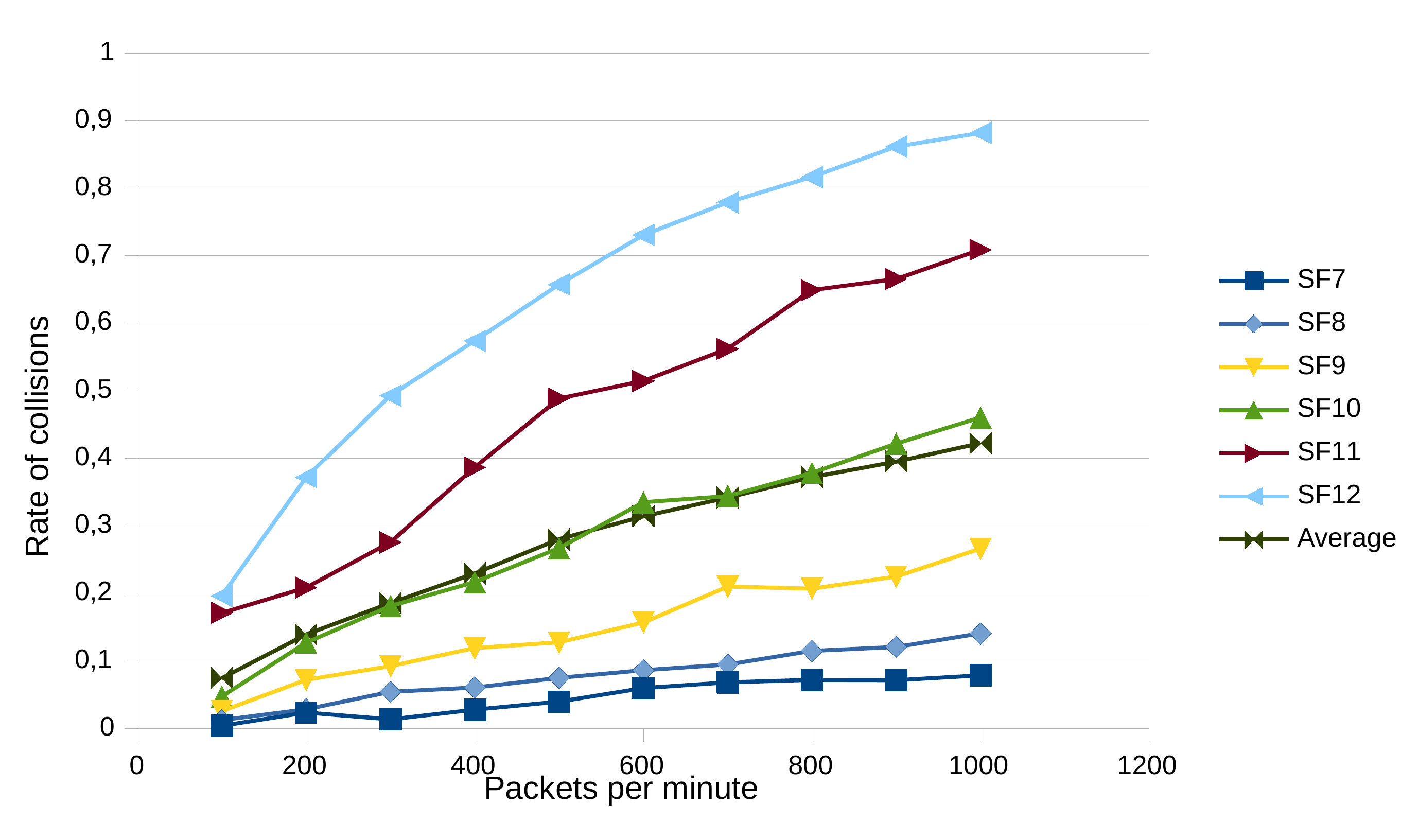}
	\caption{Rate of collisions for different SF, given $n$ packets are sent within one minute on one of the 3 EU base channels defined in the specification.}
	\label{fig:collision}
\end{figure}

Using this set-up, the airtime was calculated using Eq. (\ref{tpac}). In the simulation, whenever a frame arrived at a gateway within the same interval, SF, and channel, as any other frame, both packets were marked as colliding. The resulting number of expected collisions is shown in Fig. \ref{fig:collision}.

One may observe that a high number of collisions is to be expected once the packet rate increases. This effect can be mitigated by using lower SFs or reducing the transmission power to a value sufficient to reach only one gateway.
Since all gateways are operating in half-duplex mode, no frames can be received while a packet or confirmation is being sent. Therefore, a communication channel will saturate quickly if many nodes ask for confirmation of frames, or when a high number of down-link packets are scheduled.

\subsection{Confirmed and Down-link Frames}

Our data-set provides the number of packets sent from and received at a gateway, which allows us to estimate the ratio of received to sent packets at 0.01095. Although this ratio includes OTAA activations and a small number of down-link frames, we assume the amount of confirmed packets to be 1\%.

We extended the previously described simulation to incorporate packet confirmations, by estimating the time a gateway needs to send a confirmation; we assume the confirmation to be a message without a payload that has the ACK bit set to 1 and which has a length (corresponding to the header length) of 13 bytes. Since all gateways operate in a half-duplex mode, no frames can be received at any channel or spreading factor, while a gateway is responding with a confirmation on the same channel and spreading factor it received the frame on.

\begin{figure}[ht]
	\centering
		\includegraphics[width=1.00\linewidth]{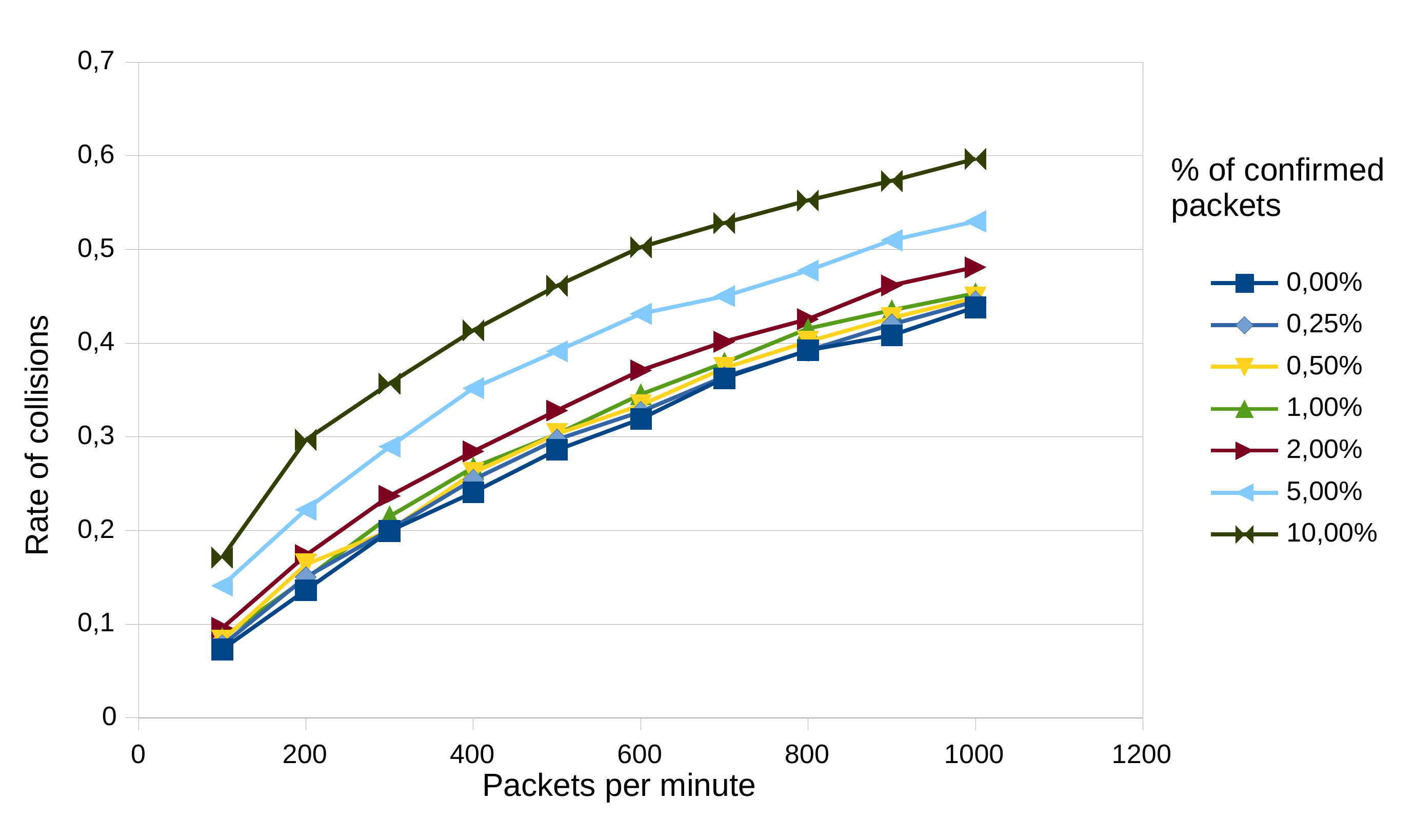}
	\caption{Rate of collisions, given $n$ packets are sent within one minute and a certain \% of packets requested confirmation.}
	\label{fig:confirmedcollision}
\end{figure}

As shown in Fig. \ref{fig:confirmedcollision}, the average packet-loss increases with the percentage of confirmed packets sent. The figure depicts the average number of expected collisions between frames on all SFs, including the non-received frames, due to the gateway acknowledging. The packet-loss on higher spreading factors (SF11 and SF12) lies above the plotted lines in Fig. \ref{fig:confirmedcollision}, whereas the lower ones (SF7, SF8 and SF9) are slightly below. Like in Fig. \ref{fig:collision}, SF10 matches the average quite well.

Given that we observed that 1\% of all packets in our data-set were requesting a confirmation, these simulations indicate that users operating LoRaWAN nodes should reduce the usage of confirmed packets, but also down-links, as much as possible. Moreover, we did not simulate re-transmissions nor other interference in the band, which would aggravate the situation. 

Another consequence of a high number of confirmed packets is that gateways, which should follow the specification as well, will quickly violate the duty-cycle regulation. Using our simulator, we estimated the total time a gateway spends on transmitting the confirmations. A configuration of 0.5\% of confirmed frames and 700 packets per minute will force the gateway to violate the regulation by using 1.28\% of the total airtime. Given 1\% of confirmed frames, this problem will already occur at a rate of 200 packets per minute. If more than 2\% of all packets request confirmations, the gateway will violate the regulation in all tested cases.

As LoRa transceivers are able to receive data, simple packet forwarders, also called single-channel gateways can be built. These are transceivers that are able to sequentially receive frames on one channel and one spreading factor, but which do not support sending data themselves. We note that in our data-set, out of the total number of 691 gateways, roughly 187 are based on the Raspberry Pi or ESP8266 platforms. From those, 97 never sent a packet to a node, and all frames were received at the same channel, which suggests that most of these 97 are indeed packet forwarders. Not being able to transmit frames means that a user has to use ABP and configure his/her device to use only one channel and SF; a fact that limits the number of packets per day and the amount of data. However, given the observation of high packet-loss, once many frames are ``in the air,'' single-channel gateways might be a cheap solution to mitigate the problem. However, network operators need to detect their presence, as to not schedule down-link packets via such packet forwarders.

\section{Related Work\label{Sec_relatedWork}}

Vangelista \textit{et al.} \cite{Vangelista2015} present LoRa as ``one of the most promising technologies for the wide-area IoT''
and mention that LoRa exhibits certain advantages over the LPWAN technologies Sigfox\texttrademark, Weightless\texttrademark, and On-Ramp Wireless. The robust chirp signal modulation and the low energy usage in combination with the low cost of end-devices together with the fact that the LoRa Alliance is also actively marketing and pushing interoperability aspects, makes LoRaWAN an interesting choice among available LPWAN technologies.

In \cite{centenaro2015long}, Centenaro \textit{et al.} provide an overview of the LPWAN paradigm in the context of smart-city scenarios. 
The authors also test the coverage of a LoRaWAN gateway in a city in Italy, by using a single base-station without antenna gain. The covered area had a diameter of 1.2 km.

The expected coverage of LPWANs and especially LoRa was also analyzed by Pet\"aj\"aj\"arvi \textit{et al.} \cite{Petajajarvi15}, who conducted measurements in Finland. Using a single base-station with an antenna gain of 2 dBi and configuring the nodes to send packets at SF12 using 14 dBm of transmit power, connectivity within a 5 km range in urban environments and 15 km in open space were found to result in packet-loss ratios smaller than 30\%. Measurements conducted by sending packets from a node mounted to a boat revealed that packets can be sent over a distance of almost 30 km.
Pet\"aj\"aj\"arvi \textit{et al.} \cite{Petajajarvi16-1} also tested the usage of LoRa in indoor environments. The results show that very low packet-loss is to be expected with only one base-station to cover an average university campus.

Bor \textit{et al.} \cite{bor2016lora} conducted experiments using multiple nodes transmitting data using LoRa. Experiments were conducted in which two devices sent packets at different power levels, but the same spreading factor, to estimate the influence of concurrent transmissions. Additionally, a new media access control (MAC), LoRaBlink, was developed to enable direct connection of nodes without using LoRaWAN.

Contrary to the above-mentioned work, we have not confined to a single base-station, but we have provided extensive measurements based on the large-scale ``The Things Networks'' network and for a duration of 8 months. 

\section{Conclusion\label{Sec_conclusion}}

In this paper, we have presented measurements from a real-live, large-scale, LoRaWAN network, along with statistics describing the use of LoRaWAN in practice, based on a data-set containing all packets, sent by devices using generic keys, within a time-frame of 8 months. Additionally, we have developed and used a simulator to study possible obstacles that might arise under heavy load of gateways.  

In order to maximize the utilization of LoRaWAN networks, certain parameters and effects should be known to the user. We have observed, for example, that not all available channels provided by a network operator are used evenly, which leads to increased packet-loss. The reason probably lies in the fact that LoRa devices are shipped with default settings complying with the LoRaWan specification. 
Although the issue will not occur if devices register via ``over the air activation,'' presently that method of connecting is only used by a limited set of devices. Moreover, we believe that, especially in mobile environments, nodes will remain to use ``activation by personalization,'' for which the channel list has to be programmed by the user. 

Since the licensed free bands are a shared good, devices should be configured in such a way that they reach a minimum amount of gateways. We also recommend using the lowest possible spreading factor, rather than limiting the output power, to reduce the needed airtime for packet transmission, to allow optimal usage of the access medium and reduce the probability of packet-loss due to collisions. 
This could be achieved by using the adaptive data rate (ADR) setting, which means a node will try to find optimal settings when transmitting the first few packets. However, for mobile scenarios we propose the contrary, not to use ADR, as it would imply quite some overhead in terms of transmitted packets, especially in urban environments, because of constant retries bloating the used airtime of a node.

As all gateways available today operate in half-duplex mode, no frames will be received while (re)transmitting or confirming data. 
To ensure a fair usage of LoRaWAN, one should aim to reduce the number of packets sent from a gateway to a node. 
Although a base-station could theoretically penalize duty-cycle violations of nodes, it is unclear how the restriction can be enforced at the gateway side, especially when multiple nodes, each respecting the regulations, send confirmed frames.

\section*{Acknowledgment}
We thank Paul Dekkers from SURFnet who provided us with a LoRa gateway and in that way kick-started this work.

\bibliographystyle{plain}
\bibliography{references}

\end{document}